\documentclass[10pt,a4paper]{article}

\newcommand{\sect}{\setcounter{equation}{0}\section}

\usepackage{theorem}
\theoremheaderfont{\sc}
\newtheorem{Thm}{Theorem}[section]
\newtheorem{Prop}{Proposition}[section]
\usepackage{t1enc}
\usepackage{amsmath}
\usepackage{amssymb}
\usepackage{enumerate}
\begin{document} 

\title{On Killing vectors in initial value problems for asymptotically
flat space-times}

\author{János Kánnár\footnote{E-mail:kannar@rmki.kfki.hu}\\
MTA KFKI \\
Részecske és Magfizikai Kutatóintézet \\
Budapest, Pf. 49, \\
1525, Hungary}

\maketitle

\begin{abstract}
\noindent
The existence of symmetries in asymptotically flat space-times are
studied from the point of view of initial value problems. General
necessary and sufficient (implicit) conditions are given for the
existence of Killing vector fields in the asymptotic characteristic
and in the hyperboloidal initial value problem (both of them are
formulated on the conformally compactified space-time manifold).
\end{abstract}

\sect{Introduction}

The most convenient way of considering the far fields of isolated
gravitational systems is to use the conformal technique introduced by
Penrose (see in \cite{penrose:conftech}). In this setting one works on
the conformally extended space-time manifold where points at infinity
(with respect to the physical metric) are glued to the physical
space-time manifold, i.e on this extended, unphysical space-time
manifold they are represented by regular points. This means that one
works on finite regions of the unphysical space-time, where one can
use all the tools of standard, local differential geometry to perform
calculations, so one avoids the determination of limits at infinity.
Of course, not all type of space-times admit the construction of
conformal infinities, those where the conformal extension can be
performed are called asymptotically simple. Asymptotically flat are
those asymptotically simple space times, where the cosmological
constant vanishes. Space-times representing isolated gravitational
systems are supposed to be asymptotically flat.

Several well-defined initial value problems can be formulated on the
extended, unphysical space-time manifold for asymptotically flat
space-times, e.g. the following initial value problems were studied
extensively in the literature.

\begin{itemize}
\item In the asymptotic characteristic initial value problem the data
are given on past null infinity ${\cal{J}}^-$ and on an incoming null
hypersurface ${\cal{N}}$ which intersects ${\cal{J}}^-$ in a
space-like surface ${\cal{Z}}$ diffeomorphic to ${\cal{S}}^2$ (the
problem could be analogously formulated for future null infinity
${\cal{J}}^+$ with an intersecting, outgoing null hypersurface, as
well).

\item In the hyperboloidal initial value problem the data are given on
a (3-dimensional) space-like hypersurface ${\cal{S}}$ intersecting
future null infinity ${\cal{J}}^+$ in a space-like surface ${\cal{Z}}$
diffeomorphic to ${\cal{S}}^2$. The term ``hyperboloidal'' comes from
the fact that the physical metric on ${\cal{S}}$ behaves near
${\cal{Z}}$ like that of a space with constant negative curvature.
This problem is not time symmetric in the sense that the Cauchy
development of a hyperboloidal hypersurface intersecting past null
infinity ${\cal{J}}^-$, in opposition to the problem formulated with
respect to ${\cal{J}}^+$, does not extends up to null infinity.
\end{itemize}
These initial value problems are solved, i.e. there are proved
uniqueness and existence theorems for these cases (see in
\cite{friedrich:acivp3},\cite{kannar:acivp},\cite{friedrich:hyperboloidal},
a review can be found in \cite{friedrich:newman}).

\begin{itemize}
\item The standard Cauchy problem is still not solved completely, it
is not known what kind of asymptotic conditions are necessary to be
imposed on the initial data in order to get smooth null infinity in
the time evolution, i.e. to get asymptotically flat space-time in the
sense defined by Penrose (the state of the research is reviewed in
\cite{friedrich:gr15}).
\end{itemize}

Working on the unphysical space-time provides some advantages even for
numerical calculations, because the evolution can be calculated over a
finite grid covering a conformally compactified initial space-like
hypersurface (some recent results can be found in Ref.
\cite{frauendiener:num1,frauendiener:num2,frauendiener:num3,hubner:h1,hubner:h2,hubner:h3,hubner:h4,hubner:h5}).

In this paper we want to formulate initial value problems for Killing
vector fields in asymptotically flat space-times. Our results will be
applicable for both of the hyperboloidal and the asymptotic
characteristic initial value problems. We will formulate general
(implicit) conditions on the initial data, which guarantee the
existence of Killing vector fields in the time evolution. Our method
is essentially the same as in \cite{racz:killing1}, where the same
problem was solved on the physical space-time manifold for all type of
the initial value problems which could be relevant there (see also
\cite{frw:killing,racz:killing2}). However, we will study here the
above introduced asymptotic characteristic and hyperboloidal initial
value problems, so we have to work in terms of the conformal
quantities on the extended, unphysical space-time manifold. The scheme
of our proofs is essentially the same as in \cite{racz:killing1},
however the corresponding calculations are much more complicated.
First we will discuss the problem in general, then in the appendix two
examples, space-times with Klein-Gordon and Maxwell fields, will be
studied in more detail.

\sect{Killing fields on the unphysical space-time}

In the following we will use the notation that quantities which are
defined with respect to the physical space-time manifold
$(\tilde{M},\tilde{g}_{ab})$ wear a tilde, and the indices of
such (tensorial) quantities will be raced and lowered by the physical
metric $(\tilde{g}_{ab},\tilde{g}^{ab})$.

Let $\tilde{\eta}^a$ be a Killing vector field on the physical
space-time manifold $(\tilde{M},\tilde{g}_{ab})$, i.e. which satisfies
the equation ${\cal{L}}_{\tilde\eta}\tilde{g}_{ab}=0$.  From now on
${\cal{L}}$ will denote the Lie derivative of the corresponding
quantity (in the previous case it was taken with respect to the vector
field $\tilde\eta^a$). The vector field $\tilde{\eta}^a$ has a unique,
smooth extension $\eta^a$ (with $\tilde\eta^a=\eta^a|_{\tilde{M}}$) to
the conformal space-time manifold $(M,g_{ab},\Omega)$, where $M$,
$\Omega$ and $g_{ab}=\Omega^2\tilde{g}_{ab}$ denote the extended,
unphysical space-time, the conformal factor (which vanishes on null
infinity) and the unphysical metric, respectively. The vector field
$\eta^a$ is tangential to null infinity $\cal{J}$ (the symbol
$\cal{J}$ will denote in this paper both of future and past null
infinity, i.e. ${\cal{J}}^+$ and ${\cal{J}}^-$, respectively),
i.e. $\eta(\Omega)|_{\cal{J}}=0$ is satisfied \cite{geroch:astr}.  On
$(M,g_{ab},\Omega)$ the vector field $\eta^a$ is a conformal Killing
vector field, i.e. equation
\begin{equation}
\begin{array}{ccc}
{\cal{L}}_{\eta}g_{ab}= \nabla_a\eta_b +\nabla_b\eta_a =
\eta(\omega)\,g_{ab} & \mathrm{with} & \omega=\mathrm{ln}(\Omega^2)
\label{conf_kill_eq}
\end{array}
\end{equation}
is satisfied,\footnote{It is worth emphasizing that $\eta(\omega)$ is
a regular expression even on null infinity, where $\Omega$ vanishes.}
where $\nabla_a$ denotes the Levi-Civita differential operator
corresponding to the conformal metric $g_{ab}$.  Substituting equation
(\ref{conf_kill_eq}) into the definition of the curvature tensor
$\nabla_a\nabla_b\eta_c-\nabla_b\nabla_a\eta_c={R_{abc}}^f\eta_f$,
then contracting with the unphysical metric $g^{ab}$, we get the
equation
\begin{equation}
\square\eta_a+\nabla_a\eta(\omega)+{R_a}^{f}\eta_f=0,
\label{eta_evol_eq}
\end{equation}
where we introduced the D'Alambert operator
$\square:=\nabla_f\nabla^f$. The above equation is satisfied by all
conformal Killing vector fields. Moreover, equation
(\ref{eta_evol_eq}) is a linear wave equation, so prescribing suitable
initial conditions we can formulate well defined initial value
problems for an arbitrary vector field $\eta_a$ satisfying the above
equation. We will show in the following that vector fields satisfying
equation (\ref{eta_evol_eq}), provided additionally with appropriate
initial data, are indeed satisfying equation (\ref{conf_kill_eq}) on
the unphysical space-time, i.e. they are proper Killing vector fields
in the physical space-time.

Differentiating equation (\ref{eta_evol_eq}) we get after some algebra
the expression
\begin{equation}
0=\square\nabla_a\eta_b+(\nabla_a{R_b}^f-\nabla_b{R_a}^f+\nabla^fR_{ab})\eta_f
+2\,R_{a\;b}^{\;\;e\;f}\nabla_e\eta_f-{R_a}^f\nabla_f\eta_b 
+{R_b}^f\nabla_a\eta_f+\nabla_a\nabla_b\eta(\omega).
\label{box_grad_eta}
\end{equation}
Introducing the tensor field
\begin{equation}
C_{ab}=\nabla_a\eta_b+\nabla_b\eta_a-\eta(\omega)g_{ab}
\label{C_def}
\end{equation}
equation (\ref{box_grad_eta}) implies the expression
\begin{equation}
0=\square{C}_{ab}+2\,R_{a\;b}^{\;\;e\;f}C_{ef}-{R_a}^fC_{fb}-{R_b}^fC_{fa}
+g_{ab}\square\eta(\omega)+2\,{\cal{L}}_{\eta}R_{ab}
+2\,\nabla_a\nabla_b\eta(\omega).
\label{C_eq_1}
\end{equation}
The last term of the previous equation can be rewritten as
\begin{equation}
2\,\nabla_a\nabla_b\eta(\omega)=
(\nabla_aC_{bf}+\nabla_bC_{af}-\nabla_fC_{ab})\nabla^f\omega
+{\cal{L}}_{\eta}(\nabla_a\omega\nabla_b\omega+2\,\nabla_a\nabla_b\omega) 
-g_{ab}(\nabla^f\omega){\cal{L}}_{\eta}\nabla_f\omega, 
\label{nn_eta_omega}
\end{equation}
while for $\square\eta(\omega)$ we can derive the equation
\begin{equation}
\square\eta(\omega)= C_{ab}\nabla^a\nabla^b\omega
-\frac{1}{2}C_{ab}(\nabla^a\omega)\nabla^b\omega
+\frac{1}{3}\eta(\omega)(\tilde{R}-R)
+\frac{1}{3}{\cal{L}}_{\eta}(\tilde{R}-R),
\label{box_eta_omega}
\end{equation}
where we introduced $\tilde{R}=g^{ef}\tilde{R}_{ef}$, while $R$
denotes the curvature scalar of the unphysical space-time. Deriving
the above equations we used several times that $\eta_a$ is a solution
of the wave equation (\ref{eta_evol_eq}).  Substituting the formulae
(\ref{nn_eta_omega}) and (\ref{box_eta_omega}) into (\ref{C_eq_1}) and
utilizing that
\begin{equation}
2\,{\cal{L}}_{\eta}(R_{ab}+\nabla_a\nabla_b\omega)=
{\cal{L}}_{\eta}\bigl\{2\,\tilde{R}_{ab} 
-\frac{1}{3}g_{ab}(\tilde{R}-R)
+\frac{1}{2}g_{ab}(\nabla_f\omega)\nabla^f\omega
-(\nabla_a\omega)\nabla_b\omega\bigr\},
\end{equation} 
where we have used the conformal transformation formula 
\begin{equation}
R_{ab}=\tilde{R}_{ab}
+\frac{3}{\Omega^2}g_{ab}(\nabla_f\Omega)\nabla^f\Omega
-\frac{1}{\Omega}\bigl\{2\,\nabla_a\nabla_b\Omega
+g_{ab}\nabla_f\nabla^f\Omega\bigr\}
\end{equation}
for the Ricci tensor, we arrive at our evolution equation
\begin{equation}
\begin{array}{l@{}l}
0= &\square C_{ab} +2\,R_{a\;b}^{\;\;e\;f}C_{ef} -{R_a}^fC_{fb}
-{R_b}^fC_{fa}
+\bigl\{\nabla_aC_{bf}+\nabla_bC_{af}-\nabla_fC_{ab}\bigr\}\nabla^f\omega 
\\[6pt]
&+\bigl\{\frac{1}{2}(\nabla_f\omega)\nabla^f\omega 
-\frac{1}{3}(\tilde{R}-R)\bigr\}C_{ab}
-g_{ab}\bigl\{(\nabla^e\omega)\nabla^f\omega 
-\nabla^e\nabla^f\omega\bigr\}C_{ef}
+2{\cal{L}}_{\eta}\tilde{R}_{ab}
\label{C_eq_2}
\end{array}
\end{equation}  
for the tensor field $C_{ab}$.

\sect{Vacuum space-times}

Equation (\ref{C_eq_2}) is a second order, linear, hyperbolic partial
differential equation for the tensor field $C_{ab}$, in vacuum
($\tilde{R}_{ab}=0$) it is additionally homogeneous.  This means that
the following assertion is just a simple consequence of the general
existence and uniqueness theorems for wave equations
(cf. Ref. \cite{taylor:book}).
\begin{Thm}
Let $(M,g_{ab},\Omega)$ denote some conformally compactified,
asymptotically flat vacuum space-time. If the vector field $\eta_a$ is
a nontrivial solution of the evolution equation (\ref{eta_evol_eq}),
furthermore the tensor field $C_{ab}$ vanishes on the initial surfaces
(hypersurface) of the considered asymptotic characteristic
(hyperboloidal) initial value problem, then
$\tilde{\eta}^a=\eta^a|_{\tilde{M}}$ is a Killing vector field on the
considered region of the physical space-time.
\end{Thm}

It is worth noting that the regularity of the principal part of
(\ref{C_eq_2}) allows the use of the standard energy estimate methods
for proving the uniqueness of the $\{C_{ab}\equiv{0}\}$ (i.e. $\eta^a$
is a conformal Killing vector) solution.

Now we turn to the more general case where also matter fields are
present in the space-time. First we will derive some general results,
then finally we discuss the cases of massless scalar and
electro-magnetic field.

\sect{Space-times with matter fields}

We start with some general assumption on the matter fields admitted in
space-times which will be discussed in our following studies. We will
suppose that the energy-impulse tensor has the structure
\begin{equation}
\tilde{T}_{ab}=\tilde{T}_{ab}(\tilde{\Phi}^{(i)}_A,
\tilde{\nabla}_e\tilde{\Phi}^{(i)}_A, \tilde{g}_{ef}),
\label{en_imp}
\end{equation}
i.e. it depends on some matter fields
$\tilde{\Phi}^{(i)}_A\equiv\tilde{\Phi}^{(i)}_{a_1{\dots}a_n}$ (where
capital indexes like $''A,B,\dots''$ are multi indexes denoting a
collection $''a_1a_2\dots''$ of covariant indexes, while $''i''$ is
just to label the several matter fields), on their first covariant
derivatives and on the physical metric. The fields
$\tilde{\Phi}^{(i)}_A$ are supposed to have regular limits at null
infinity $\cal{J}$, so they can be extended smoothly to well-defined
tensor fields $\Phi^{(i)}_A$ on the unphysical space-time where
$\tilde{\Phi}^{(i)}_A=\Phi^{(i)}_A|_{\tilde{M}}$ is
satisfied. Supposing that the Einstein equation holds, we get an
expression similar to (\ref{en_imp}) for the physical Ricci
tensor\footnote{It is worth remarking, that we could have started,
just like in \cite{racz:killing1}, with imposing the conditions
(\ref{ricci_tensor}) and (\ref{field_evol_eq}), the analysis itself is
independent of the exact form of the Einstein equation.}
\begin{equation}
\tilde{R}_{ab}=\tilde{R}_{ab}(\tilde{\Phi}^{(i)}_A,
\tilde{\nabla}_e\tilde{\Phi}^{(i)}_A, \tilde{g}_{ef}).
\label{ricci_tensor}
\end{equation}
		
Equation (\ref{C_eq_2}) contains the Lie derivative of the physical
Ricci tensor
\begin{equation}
{\cal{L}}_{\eta}\tilde{R}_{ab}=
\displaystyle{\sum_{i}}\frac{\partial\tilde{R}_{ab}}
{\partial\tilde{\Phi}^{(i)}_A}
{\cal{L}}_{\eta}\tilde{\Phi}^{(i)}_A
+\displaystyle{\sum_{i}}\frac{\partial\tilde{R}_{ab}}
{\partial\tilde{\nabla}_e\tilde{\Phi}^{(i)}_A}
{\cal{L}}_{\eta}\tilde{\nabla}_e\tilde{\Phi}^{(i)}_A
+\frac{\partial\tilde{R}_{ab}}{\partial\tilde{g}_{ef}}
{\cal{L}}_{\eta}\tilde{g}_{ef}.
\label{ricci_lie_der_1}
\end{equation}
Like the matter fields $\tilde{\Phi}^{(i)}_A$, their Lie derivatives
${\cal{L}}_{\eta}\tilde{\Phi}^{(i)}_A$ are also well-defined tensor
fields on the whole unphysical space-time, more precisely
${\cal{L}}_{\eta}\tilde{\Phi}^{(i)}_A=
{\cal{L}}_{\eta}\Phi^{(i)}_A|_{\tilde{M}}$ is satisfied. This means
that we can omit the tildes also from the $\Phi^{(i)}_A$-s and the Lie
derivative appearing in the second term of equation
(\ref{ricci_lie_der_1}) can be rewritten as
\begin{equation}
{\cal{L}}_{\eta}(\tilde{\nabla}_e\Phi^{(i)}_{a_1{\dots}a_n})=
\tilde{\nabla}_e{\cal{L}}_{\eta}\Phi^{(i)}_{a_1{\dots}a_n} 
-\displaystyle{\sum_{j=1}^n}(\tilde{\nabla}{\cal{L}}_{\eta}\tilde{g})_{ea_jf}
\tilde{g}^{fh}\Phi^{(i)}_{a_1{\dots}h{\dots}a_n},
\label{lie_nabla_phi}
\end{equation}
where we used the abbreviation 
\begin{equation}
(\tilde{\nabla}{\cal{L}}_{\eta}\tilde{g})_{ea_jf}= 
\frac{1}{2}\left(\tilde{\nabla}_e{\cal{L}}_{\eta}\tilde{g}_{a_jf}
+\tilde{\nabla}_{a_j}{\cal{L}}_{\eta}\tilde{g}_{ef} 
-\tilde{\nabla}_f{\cal{L}}_{\eta}\tilde{g}_{ea_j}\right).
\end{equation}
Both of the above equations contain the Levi-Civita differential
operator $\tilde{\nabla}$ induced by the physical metric
$\tilde{g}_{ab}$. In order to extend these expressions into the
unphysical space-time first we have to change to the operators
$\nabla$ induced by the conformal metric $g_{ab}$, i.e. we have to
apply the conformal transformations
\begin{equation}
\begin{array}{r@{}l}
\tilde{\nabla}_e{\cal{L}}_{\eta}\Phi^{(i)}_{a_1{\dots}a_n}=
&\nabla_e{\cal{L}}_{\eta}\Phi^{(i)}_{a_1{\dots}a_n} 
+\displaystyle{\sum_{j=1}^n}
\hat{\Gamma}_{ea_j}^{\;\;\;\;f}
{\cal{L}}_{\eta}\Phi^{(i)}_{a_1{\dots}f{\dots}a_n}, \\[12pt]
\tilde{\nabla}_e{\cal{L}}_{\eta}\tilde{g}_{ab}=
&\frac{1}{\Omega^2}\nabla_eC_{ab}
+\frac{1}{\Omega^2}\left(\hat\Gamma_{ea}^{\;\;\;f}C_{fb}
+\hat\Gamma_{eb}^{\;\;\;f}C_{af} -\frac{2}{\Omega}C_{ab}\nabla_e\Omega\right),
\label{conf_trans_1}
\end{array}
\end{equation}
where we used the symbols
\begin{equation}
\hat{\Gamma}_{ab}^{\;\;\;c}=\frac{2}{\Omega}
\left(\delta_{(a}^c\nabla_{b)}\Omega
-\frac{1}{2}g_{ab}g^{cd}\nabla_d\Omega\right).
\end{equation}
We applied also the relation
\begin{equation}
{\cal{L}}_{\eta}\tilde{g}_{ab}=\frac{C_{ab}}{\Omega^2}
\label{conf_trans_2}
\end{equation}
which follows directly from the definition (\ref{C_def}) of the
tensor field $C_{ab}$. Substituting all the above-quoted formulae into
expression (\ref{ricci_lie_der_1}) we get an equation with the
structure
\begin{equation}
{\cal{L}}_{\eta}\tilde{R}_{ab}={\cal{A}}_{ab}({\cal{L}}_{\eta}\Phi^{(i)}_A)
+{\cal{B}}_{ab}(\nabla_e{\cal{L}}_{\eta}\Phi^{(i)}_A) 
+{\cal{C}}_{ab}(C_{ef}) +{\cal{D}}_{ab}(\nabla_eC_{fg}),
\label{ricci_lie_der_2}
\end{equation}
where all of ${\cal{A}}_{ab}$, ${\cal{B}}_{ab}$, ${\cal{C}}_{ab}$ and
${\cal{D}}_{ab}$ are linear, homogeneous functions in their indicated
arguments. However, as we will see later on the explicit examples,
some of these functions can contain also terms with negative powers of
the conformal factor $\Omega$ which vanishes on null infinity. This
means that also singular terms can appear on the right hand side of
(\ref{ricci_lie_der_2}).

Let us suppose that the matter fields
$\tilde\Phi^{(i)}_A=\Phi^{(i)}_A|_{\tilde{M}}$ satisfy the field
equations
\begin{equation}
\tilde{\nabla}_a\tilde{\nabla}^a \tilde\Phi^{(i)}_A =
F^{(i)}_A(\tilde\Phi^{(j)}_B, \tilde{\nabla}_e\tilde\Phi^{(j)}_B,
\tilde{g}_{ef}),
\label{field_evol_eq}
\end{equation}
where $F^{(i)}_A$ denote some smooth functions of the indicated
arguments (the matter fields are admitted to be coupled to each
other).  Calculating the Lie derivative of the previous equation, then
performing the same transformations as above, we can derive the
evolution equations
\begin{equation}
\square{\cal{L}}_{\eta}\Phi^{(i)}_A=
{\cal{E}}^{(i)}_A({\cal{L}}_{\eta}\Phi^{(j)}_B)
+{\cal{F}}^{(i)}_A(\nabla_e{\cal{L}}_{\eta}\Phi^{(j)}_B)
+{\cal{G}}^{(i)}_A(C_{ef}) +{\cal{H}}^{(i)}_A(\nabla_eC_{fg})
\label{box_lie_phi}
\end{equation}
for the Lie derivatives of the matter fields (we already performed the
conformal transformations (\ref{conf_trans_1})-(\ref{conf_trans_2}),
as well). The functions ${\cal{E}}^{(i)}_A$, ${\cal{F}}^{(i)}_A$,
${\cal{G}}^{(i)}_A$ and ${\cal{H}}^{(i)}_A$ are linear and homogeneous
in their indicated arguments. However, their regularity, like above at
(\ref{ricci_lie_der_2}), depends on the concrete physical model which
is investigated.

Equations (\ref{C_eq_2}) (in view of eq. (\ref{ricci_lie_der_2})) and
(\ref{box_lie_phi}) compose a system of linear, homogeneous wave
equations for the variables $C_{ab}$ and
${\cal{L}}_{\eta}\Phi_A^{(i)}$. The right hand sides can contain terms
which are singular on null infinity, however the principal parts are
always regular. This admits us to apply the general theorems (see in
\cite{taylor:book}) for proving the uniqueness of the
$\{C_{ab}\equiv{0}$, ${\cal{L}}_{\eta}\Phi_A^{(i)}\equiv{0}\}$
solution.  This means that the following assertion is just a simple
consequence of the general theorems.
\begin{Thm}
Let $(M,g_{ab},\Omega)$ denote some conformally compactified
asymptotically flat space-time containing some matter fields
$\Phi_A^{(i)}$ which satisfy the evolution equations
(\ref{field_evol_eq}). If the vector field $\eta_a$ is a nontrivial
solution of the wave equation (\ref{eta_evol_eq}), furthermore
${\cal{L}}_{\eta}\Phi_A^{(i)}$ and the tensor field $C_{ab}$ vanish on
the initial surfaces (hypersurface) of the considered asymptotic
characteristic (hyperboloidal) initial value problem, then
$\tilde\eta^a=\eta^a|_{\tilde{M}}$ is a Killing vector field on the
considered region of the physical space-time.
\end{Thm}

In the appendix we will consider two examples, the massless scalar and
the electro-magnetic field, in more detail. First, it is useful to
write down the actual formulae where one can observe explicitly the
nature of the singularities appearing on $\cal{J}$, caused by the
$\frac{1}{\Omega}$ terms.  Secondly, the field equations for the
electro-magnetic field are not automatically of the form
(\ref{field_evol_eq}), one have to impose suitable gauge conditions in
order to apply the above general results.

\sect{Summary}

We have derived general necessary and sufficient initial conditions
for Killing vector fields in the asymptotic characteristic and
hyperboloidal initial value problems. These conditions are implicit in
the sense that they have to be evaluated for the components of the
vector field $\eta^a$. The existence of Killing vector fields depends
on the existence of $\eta^a\neq{0}$ solutions for the conditions
derived above for the initial data. Unfortunately the evaluation of
our general implicit conditions cannot be done so general as we
treated the whole problem in this paper, e.g. the calculations are
fundamentally different for the asymptotic characteristic and
hyperboloidal initial value problems, caused by the different causal
nature of the initial manifolds. However, a future work is planed for
studying some applications of the introduced general assertions.

\vskip .5cm
\parindent 0pt
\textsc{Acknowledgments:} I would like to thank I. Rácz for reading the
manuscript.  This research was supported by the grant OTKA-D25135.
\parindent 10pt

\appendix

\sect{APPENDIX}

\subsection{Massless scalar and electro-magnetic field}

Let as consider a massless scalar field $\tilde\Phi$ on the physical
space-time manifold $(\tilde{M},\tilde{g}_{ab})$ with evolution
equation
\begin{equation}
\tilde{\nabla}_a\tilde{\nabla}^a\tilde\Phi= 0,
\end{equation}
and with energy-impulse tensor
\begin{equation}
\tilde{T}_{ab}= \tilde{\nabla}_a\tilde\Phi\tilde{\nabla}_b\tilde\Phi 
-\frac{1}{2}\tilde{g}_{ab}\tilde{\nabla}_f\tilde\Phi\tilde{\nabla}^f\tilde\Phi.
\end{equation}
The scalar field $\tilde\Phi$ is supposed to have regular limit on
null infinity, so it has a unique extension $\Phi$ (with
$\tilde\Phi=\Phi|_{\tilde{M}}$) which is regular on the whole
unphysical space-time $(M,g_{ab},\Omega)$.  Through the Einstein
equations the Lie derivative of the physical Ricci tensor with respect
to $\eta^a$ is simply
\begin{equation}
{\cal{L}}_{\eta}\tilde{R}_{ab}=
8\pi\left[(\nabla_a{\cal{L}}_{\eta}\Phi)\nabla_b\Phi 
+(\nabla_a\Phi)\nabla_b{\cal{L}}_{\eta}\Phi\right].
\label{scal_ricci_lie_der}
\end{equation}
Here we already performed the conformal transformation, the above
expression is written already in terms of the unphysical
quantities. We can recognize that (\ref{scal_ricci_lie_der}) is a
completely regular expression on the whole conformal space-time
manifold.

We can evaluate (\ref{box_lie_phi}), as well. After some longer but
straightforward calculations we arrive at the equation
\begin{equation}
0=\square{\cal{L}}_{\eta}\Phi -(\nabla^e\nabla^f\Phi) C_{ef}
-g^{ef}\nabla^h\Phi\bigl\{\nabla_eC_{fg}-\frac{1}{2}\nabla_gC_{ef}\bigr\}
-2\frac{\nabla^e\Omega}{\Omega}\bigl\{\nabla_e{\cal{L}}_{\eta}\Phi
-(\nabla^f\Phi)C_{ef}\bigr\}.
\label{scal_box_lie_phi}
\end{equation}
Equations (\ref{C_eq_2}) (in view of (\ref{scal_ricci_lie_der})) and
(\ref{scal_box_lie_phi}) constitute a linear, homogeneous system of
wave equations for the variables $C_{ab}$ and ${\cal{L}}_{\eta}\Phi$,
so the following statement is just a special case of the general
theorem formulated in the previous section.
\begin{Prop}
Let $(M,g_{ab},\Omega)$ denote some conformally compactified
asymptotically flat space-time containing some massless scalar field
in the considered region. If the vector field $\eta_a$ is a nontrivial
solution of the evolution equation (\ref{eta_evol_eq}), furthermore
${\cal{L}}_{\eta}\Phi$ and the tensor field $C_{ab}$ vanish on the
initial surfaces (hypersurface) of the considered asymptotic
characteristic (hyperboloidal) initial value problem, then
$\tilde\eta^a=\eta^a|_{\tilde{M}}$ is a Killing vector field on the
considered region of the physical space-time.
\end{Prop}

Now we turn to the study of the electro-magnetic field. The field
equations in the physical space-time are given by
\begin{equation}
(^*d\tilde{F})_a=0, \hskip 1cm \tilde{F}_{ab}=(d\tilde{A})_{ab},
\end{equation}
where $d$ denotes the exterior differential of the corresponding
quantity and the star indicates the Hodge-dual.  The vector potential
is considered as the restriction to $\tilde{M}$ of a
$\tilde{A}_a=A_a|_{\tilde{M}}$ one-form field $A_a$ which is regular
on the whole conformally extended space-time manifold.  The
electro-magnetic field equations are conformally invariant, so the
previous equations can be rewritten as
\begin{equation}
\nabla^aF _{ab}=0, \hskip 1cm F_{ab}= \nabla_aA_b -\nabla_bA_a,
\label{em_eq}
\end{equation}
where we already used the conformal Levi-Civita differential operator
induced by the unphysical metric $g_{ab}$. The energy-impulse tensor
and the physical Ricci tensor, using the Einstein equations, can be
written simply as
\begin{equation}
\tilde{T}_{ab}=F_{ae}F_{bf}\tilde{g}^{ef}
-\frac{1}{4}\tilde{g}_{ab}F_{ce}F_{df}\tilde{g}^{cd}\tilde{g}^{ef},
\hskip 1cm
\tilde{R}_{ab}=8\pi\left[F_{ae}F_{bf}\tilde{g}^{ef}
-\frac{1}{4}\tilde{g}_{ab}F_{ce}F_{df}\tilde{g}^{cd}\tilde{g}^{ef}\right],
\end{equation}
respectively. It is easy to check that the Lie derivative
(\ref{ricci_lie_der_2}) of the physical Ricci tensor now takes the
form
\begin{equation}
{\cal{L}}_{\eta}\tilde{R}_{ab}=
\Omega^2\bigl\{{\cal{A}}_{ab}({\cal{L}}_{\eta}A_e)
+{\cal{C}}_{ab}(C_{ef})\bigr\},
\end{equation}
where ${\cal{A}}_{ab}$ and ${\cal{C}}_{ab}$ are some regular
functions, homogeneous and isotropic in their indicated arguments.

The evaluation of (\ref{box_lie_phi}) requires a bit more work.
Calculating the Lie-derivative of the first equation from
(\ref{em_eq}) after some lengthy but straightforward calculation we
arrive at
\begin{equation}
\begin{array}{l@{}l}
0=&\square{\cal{L}}_{\eta}A_a
-g^{fg}\bigl\{\nabla^e\nabla_fA_a-\nabla^e\nabla_aA_f\bigr\}C_{eg}
+g^{gh}\nabla^fA_h\bigl\{\nabla_gC_{af} -\nabla_fC_{ag}\bigr\}
-{R_a}^g{\cal{L}}_{\eta}A_g \\[6pt]
&-\nabla_a\bigl\{({\cal{L}}_{\eta}+\eta(\omega))(\nabla^fA_f-A(\omega))
+g^{ef}g^{gh}(\nabla_eA_g-A_e\nabla_g\omega)C_{fh}
+(\nabla^f\omega){\cal{L}}_{\eta}A_f\bigr\}.
\label{em_box_lie_1}
\end{array}
\end{equation}
During the derivation of the previous equation we used only identity
(\ref{lie_nabla_phi}), equation (\ref{conf_trans_2}) and the evolution
equation (\ref{eta_evol_eq}) satisfied by the vector field $\eta_a$.

It is easy to check that in terms of the conformal quantities the
Lorentz gauge condition can be rewritten as
\begin{equation}
\tilde{\nabla}^f\tilde{A}_f= \Omega^2[\nabla^fA_f-A(\omega)]=0.
\end{equation}
This means that in the Lorenz gauge equation (\ref{em_box_lie_1})
takes a more simple form
\begin{equation}
\begin{array}{l@{}l}
0=&\square{\cal{L}}_{\eta}A_a
-g^{fg}\bigl\{\nabla^e\nabla_fA_a-\nabla^e\nabla_aA_f\bigl\}C_{eg}
-g^{gh}\nabla^fA_h\bigl\{\nabla_aC_{fg}+\nabla_fC_{ag}-\nabla_gC_{af}\bigr\} 
\\[6pt]
&-{R_a}^g{\cal{L}}_{\eta}A_g
-g^{fh}g^{gk}\bigl\{\nabla_a\nabla_hA_k-\nabla_aA_h\nabla_k\omega 
-A_h\nabla_a\nabla_k\omega\bigr\}C_{fg} \\[6pt]
&-(\nabla_a\nabla^f\omega){\cal{L}}_{\eta}A_f 
-(\nabla^f\omega)\nabla_a{\cal{L}}_{\eta}A_f.
\label{em_box_lie_2}
\end{array}
\end{equation}
This expression has already the structure of (\ref{box_lie_phi}).  At
this point we can argue like above, i.e. equations
(\ref{em_box_lie_2}) and (\ref{C_eq_2}) constitute a linear,
homogeneous system of wave equations with a unique
${\cal{L}}_{\eta}A_a\equiv{0}$ and $C_{ab}\equiv{0}$ solution. So the
following statement is just a special case of the general theorem of
the previous section.
\begin{Prop}
Let $(M,g_{ab},\Omega)$ denote some conformally compactified
asymptotically flat space-time containing electro-magnetic field in the
considered region. Let the vector potential $\tilde{A}_a$ is given in
the Lorentz-gauge, i.e. $\nabla^a\tilde{A}_a=0$ satisfied. If the
vector field $\eta_a$ is a nontrivial solution of the evolution
equation (\ref{eta_evol_eq}), furthermore ${\cal{L}}_{\eta}A_a$ and
the tensor field $C_{ab}$ vanish on the initial surfaces
(hypersurface) of the considered asymptotic characteristic
(hyperboloidal) initial value problem, then
$\tilde\eta_a=\eta_a|_{\tilde{M}}$ is a Killing vector field in the
considered region of the physical space-time.
\end{Prop}

\bibliography{ckvbib} 
\bibliographystyle{bibstyle}

\end{document}